\documentclass[aps,prl,twocolumn,showpacs,preprintnumbers,superscriptaddress,groupedaddress]{revtex4}
\pdfoutput=1
\usepackage{graphicx}  
\usepackage{dcolumn}   
\usepackage{bm}        
\usepackage{slashed}
\usepackage{amssymb} \usepackage{amsmath}   
 \usepackage{hyperref}
\hypersetup{colorlinks=true,linkcolor=magenta,anchorcolor=green,citecolor=cyan,filecolor=black,menucolor=black,urlcolor=black}

\usepackage[force]{feynmp-auto}

\usepackage{tikz}
 \usetikzlibrary{calc,decorations.pathreplacing}

\def\tr{\textrm{tr}}

\def\sq[#1,#2]{\left[#1\,#2\right]}
\def\an[#1,#2]{\left\langle#1\,#2\right\rangle}
\def\spab[#1,#2,#3]{\left\langle#1|#2|#3\right]}

\begin{document}
\preprint{IPHT-t18/002}
\title{General Relativity from Scattering Amplitudes}
\author{N.~E.~J.~Bjerrum-Bohr}
\author{Poul H. Damgaard}
\affiliation{Niels Bohr  International  Academy and  Discovery
Center,
The Niels Bohr Institute, Blegdamsvej 17,
DK-2100 Copenhagen \O, Denmark}
\author{Guido Festuccia}
\affiliation{Department of Physics and Astronomy, Theoretical Physics,
\AA ngstr\"omlaboratoriet, L\"agerhyddsv\"agen 1, Box 516
751 20 Uppsala, Sweden}
\author{Ludovic~Plant\'e}
\affiliation{Minist\`ere de l'\'economie et des finances, Direction
  g\'en\'erale des entreprises, France}
\author{Pierre~Vanhove}
\affiliation{Institut de Physique Th\'eorique, Universit\'e  Paris-Saclay, CEA, CNRS, F-91191 Gif-sur-Yvette Cedex, France}
\affiliation{National Research University Higher School of Economics, Russian Federation}

\date{\today}

\begin{abstract}
We outline the program to apply modern quantum field theory methods to calculate observables in classical 
general relativity through a truncation to classical terms of the
multi-graviton  two-body on-shell scattering amplitudes between massive fields.  
Since only long-distance interactions corresponding to non-analytic pieces need to be included, unitarity cuts provide substantial simplifications for both post-Newtonian and post-Minkowskian expansions.
We illustrate this quantum field theoretic approach to classical general relativity 
by computing the interaction potentials to second order in the post-Newtonian expansion, as well as the scattering functions 
for two massive objects to 
second order in the post-Minkowskian expansion. 
We also derive an all-order exact result for gravitational light-by-light scattering.
\end{abstract}
\maketitle

Today it is universally accepted that classical general relativity can
be understood as the $\hbar \to 0$ limit of a quantum mechanical path
integral with an action that, minimally, includes the Einstein-Hilbert
term. It describes gravitational interactions in terms of exchanges
and interactions of spin-2 gravitons with themselves (and with matter)~\cite{Feynman:1963ax,DeWitt:1967yk}. The language of effective field
theory encompasses this viewpoint, and it shows that a large-distance
quantum field theoretic description of gravity is well defined order
by order in a derivative expansion~\cite{Weinberg,Donoghue:1993eb}.
Quantum mechanics thus teaches us that we should expect classical
general relativity to be augmented by higher-derivative terms. More
remarkably, what would ordinarily be a quantum mechanical loop
expansion contains pieces at arbitrarily high order that are entirely
classical~\cite{Iwasaki:1971vb,Holstein:2004dn}.  A subtle cancellation of
factors of $\hbar$ is at work here. This leads to the radical
conclusion that one can define classical general relativity
perturbatively through the loop expansion. Then $\hbar$ plays a role
only at intermediary steps, a dimensionful regulator that is unrelated
to the classical physics the path integral describes.

For the loop expansion, central tools have been the unitarity methods~\cite{Bern:1994zx} which reproduce those parts of loop
amplitudes that are ``cut constructable'', $i.e.$ all non-analytic
terms of the amplitudes. This amounts to an enormous simplification,
and most of today's amplitude computations for the Standard Model of
particle physics would not have been possible without this method. In
classical gravity, the long-distance terms we seek are precisely of
such non-analytic kind, being functions of the dimensionless ratio
$m/\sqrt{-q^2}$, where $m$ is a massive probe, and $q^\mu$ describes a
suitably defined momentum transfer~\cite{Donoghue:1993eb}. This leads
to the proposal that these modern methods be used to
compute post-Newtonian and post-Minkowskian
perturbation theory of general relativity for astrophysical processes such as
binary mergers. This has acquired new urgency due to the recent observations
of gravitational waves emitted during such inspirals.

While the framework for classical general relativity as described
above would involve all possible interaction terms in the Lagrangian,
ordered according to a derivative expansion, one can always choose to
retain only the Einstein-Hilbert action. Quantum mechanically this is
inconsistent, but for the purpose of extracting only classical results
from that action, it is a perfectly valid truncation. This scheme
relies on a separation of the long-distance (infrared) and
short-distance (ultraviolet) contributions
in the scattering amplitudes in quantum field theory. We will follow
that strategy here, but one may apply the same amplitude methods to
actions that contain, already at the classical level,
higher-derivative terms as well.  In the future, this may be used to
put better observational bounds on such new couplings.

In~\cite{Damour:2016gwp} Damour proposed a new approach for converting
classical scattering amplitudes into the effective-one-body
  Hamiltonian  of two gravitationally interacting bodies. In this work
  we take a different route and 
  we show how scattering amplitude methods, which build on the probabilistic
nature of quantum mechanics, may be used to derive  classical
results in gravity. We show how tree-level massless emission from massive 
classical sources arises from quantum multiloop amplitudes, thus providing an
all-order argument extending the original observations in~\cite{Holstein:2004dn}. 
We apply this method to derive the scattering angle between two masses to second
post-Minkowskian order using the eikonal method.

 We start with the Einstein-Hilbert action coupled to a scalar field $\phi$
\begin{equation}
{\cal S}\!=\!\int d^4x \sqrt{-g}\left[{1\over16\pi G} R\! +\! \frac12
  g^{\mu\nu}\partial_\mu \phi\partial_\nu\phi\!-\!{m^2\over2}\phi^2\right]\,.
\end{equation} 
Here $R$ is the curvature and $g_{\mu\nu}$ is the metric, defined as
the sum of a flat Minkowski component $\eta_{\mu\nu}$ and a
perturbation $\kappa h_{\mu\nu}$ with $\kappa \equiv \sqrt{32\pi G}$. It is coupled to the scalar stress-energy
tensor $T_{\mu\nu} \equiv \partial_\mu \phi\, \partial_\nu \phi 
- \frac{\eta_{\mu\nu}}{2}\left(\partial^\rho\phi\partial_\rho\phi-m^2 \phi^2\right)$.

Scalar triangle integrals~\cite{Bern:1992em} are what reduces the
one-loop two-graviton scattering amplitude to classical general
relativity~\cite{Donoghue:1993eb,Neill:2013wsa,Bjerrum-Bohr:2013bxa,Vaidya:2014kza} in four dimensions.
For the long-distance contributions these are the integrals that produce the tree-like structures
one intuitively associates with classical general relativity. To see this,
consider first the triangle integral of one massive and two massless propagators, 
\begin{multline}
 I_\triangleright(p_1,q)=\begin{gathered}\begin{fmffile}{triangleleft1}
  \begin{fmfgraph}(100,50)
      \fmfleftn{i}{2}
      \fmfrightn{o}{1}
\fmf{plain}{i1,v1}
\fmf{plain}{i2,v2}
\fmf{dbl_wiggly}{o1,v3}
\fmf{plain,tension=.1}{v1,v2}
\fmf{dbl_wiggly,tension=.3}{v3,v1}
\fmf{dbl_wiggly,tension=.3}{v3,v2}
\end{fmfgraph}
\end{fmffile}\end{gathered}\cr
\!\!\!=\int {d^4\ell\over (2\pi)^4} {1\over \ell^2+i\epsilon}{1\over (\ell+q)^2+i\epsilon} {1\over (\ell+p_1)^2-m_1^2+i\epsilon}  \,,
\end{multline}
with $p_1=(E,\vec q/2)$, $p_2=(E,-\vec q/2)$ and $q\equiv p_1-p_1'=(0,\vec
q)$ and $E=\sqrt{m_1^2+\vec q^2/4}$, and we work with the mostly
negative metric $(+---)$.
The curly lines are for massless fields and
the left solid line is for a particle of incoming momentum
$p_1$, outgoing momentum $p_1'$ and mass $p_1^2={p_1'}^2=m_1^2$.

In the large mass approximation we focus on the region $|\vec\ell|\ll m_1$
we have $(\ell+p_1)^2-m_1^2=\ell^2+2\ell\cdot p_1\simeq 2m_1 \ell_0$ therefore the integral reduces in that limit to 
\begin{equation}\label{e:Tril0}
{1\over 2m_1}\!\int {d^4\ell\over (2\pi)^4}\! {1\over \ell^2+i\epsilon}{1\over
  (\ell+q)^2+i\epsilon} {1\over \ell_0+i\epsilon}  \,.
\end{equation}
We perform the $\ell_0$ integral by closing the contour of integration
in the upper half-plane to get
\begin{equation}\label{e:Triresult}
\int_{|\vec\ell|\ll m} {d^3\vec\ell\over(2\pi)^3} {i\over 4m} {1\over \vec\ell^2}{1\over
    (\vec\ell+q)^2}=-{i\over 32m |\vec q|}\,.
\end{equation}
This result can be obtained by performing the large mass expansion of
the exact expression for the triangle integral as shown in the Appendix.

In~\eqref{e:Triresult} we recognize the three-dimensional
integral of two static sources localized at different positions,
represented as shaded blobs, and emitting massless fields
\begin{equation}
\int {d^3\vec\ell\over(2\pi)^3} {1\over \vec\ell^2}{1\over
    (\vec\ell+q)^2}\longleftrightarrow
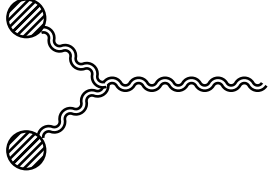
\begin{gathered}\begin{fmffile}{classicaltree1}
    \begin{fmfgraph*}(100,50)
       \fmfleftn{i}{2}
        \fmfrightn  {o}{1}
\fmf{dbl_wiggly}{i1,v1}
\fmf{dbl_wiggly}{i2,v1}
\fmf{dbl_wiggly}{v1,o1}
\fmfblob{.5cm}{i1}
\fmfblob{.5cm}{i2}
\end{fmfgraph*}
\end{fmffile}
\end{gathered}
\end{equation}
Below we show how this allows us to recover the first post-Newtonian correction to
the Schwarzschild metric from quantum loops. We now explain how the classical part emerges from
higher-loop triangle graphs, starting with two-loop triangles 
\begin{align}
& I_{\triangleright\triangleright (1)}(p_1,q)=
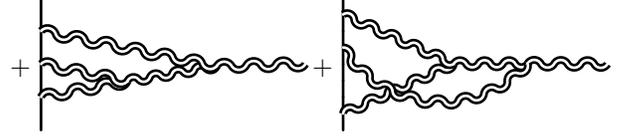
\begin{gathered}
\begin{fmffile}{twolooptriangle4pt12}
 \begin{fmfgraph}(100,50)
\fmfstraight
      \fmfleftn{i}{2}
      \fmfrightn{o}{1}
\fmf{plain,tension=10}{i1,v1,v4,v2,i2}
\fmffreeze
\fmf{dbl_wiggly,tension=3}{v3,v5,v2}
\fmf{dbl_wiggly}{v3,v1}
\fmf{dbl_wiggly,tension=.1}{v5,v4}
\fmf{dbl_wiggly,tension=5}{v3,o1}
\end{fmfgraph}
\end{fmffile}
\end{gathered}\\
\nonumber&+
\begin{gathered}
\begin{fmffile}{twolooptriangle4pt21}
  \begin{fmfgraph}(100,50)
\fmfstraight
      \fmfleftn{i}{2}
      \fmfrightn{o}{1}
\fmf{plain,tension=10}{i1,v1,v4,v2,i2}
\fmffreeze
\fmf{dbl_wiggly,tension=3}{v3,v5,v1}
\fmf{dbl_wiggly}{v3,v2}
\fmf{dbl_wiggly,tension=.1}{v5,v4}
\fmf{dbl_wiggly,tension=5}{o1,v3}
\end{fmfgraph}
\end{fmffile}
\end{gathered}
+
\begin{gathered}
\begin{fmffile}{twolooptriangle3}
  \begin{fmfgraph}(100,50)
\fmfstraight
      \fmfleftn{i}{2}
      \fmfrightn{o}{1}
\fmf{plain}{i1,v1,vph1a,vph1b,vph1c,v4,vph2,v2,i2}
\fmffreeze
\fmf{dbl_wiggly}{v3,v1}
\fmf{dbl_wiggly}{v3,v2}
\fmf{dbl_wiggly,tension=3}{o1,v5,v3}
\fmf{dbl_wiggly,left=.5,tension=.1}{v5,v4}
\end{fmfgraph}
\end{fmffile}
\end{gathered}
\end{align}

In  the large mass limit  $|\vec\ell_i|\ll m_1$ for $i=1,2,3$ and approximating
$(\ell_i+p_1)^2-m_1^2\simeq 2\ell_i\cdot p_1\simeq 2m_1\ell_i^0$ the
integral reduces in that limit to
\begin{multline}
  I_{\triangleright\triangleright (1)}(p_1,q) =  -\int \prod_{i=1}^3 {d^4\ell_i\over (2\pi)^4}
 {1\over \ell_i^2+i\epsilon}\cr
\times{(2\pi)^3\delta^{(3)}(\sum_{i=1}^3 \vec\ell_i+\vec q)\over
    (\ell_1+q)^2+i\epsilon}
\times2\pi
\delta(\ell_1^0+\ell_2^0+\ell_3^0)\cr
\times\Big({1\over
    2m_1\ell_1^0+i\epsilon}{1\over
    2m_1\ell_2^0-i\epsilon}
+{1\over
   2m_1 \ell_3^0+i\epsilon}{1\over
   2m_1 \ell_1^0-i\epsilon}
\cr+{1\over
    2m_1\ell_3^0+i\epsilon}{1\over
    2m_1\ell_2^0-i\epsilon}
\Big)\,.
\end{multline}
We note that 
\begin{multline}\label{e:identity}
 \delta(\ell_1^0+\ell_2^0+\ell_3^0)\Big( {1\over
    2m_1\ell_1^0+i\epsilon}{1\over
  2m_1  \ell_2^0-i\epsilon}
\cr+{1\over
   2m_1 \ell_3^0+i\epsilon}{1\over
    2m_1\ell_1^0-i\epsilon}
+{1\over
    2m_1\ell_3^0+i\epsilon}{1\over
   2m_1 \ell_2^0-i\epsilon}\Big)\cr=0+O(\epsilon)\,,
\end{multline}
so that only the 
$\ell_0$ 
residue at $2m_1\ell^0=\pm i\epsilon$ contributes, giving 
\begin{equation}\label{e:Tri2loop1}
  I_{\triangleright\triangleright (1)}(p_1,q) =  {i\over 4m_1^2}\!\!\! \int\!\!\! {d^3\vec\ell_1\over (2\pi)^3}\!\! {d^3\vec\ell_2\over (2\pi)^3}
 {1\over \vec\ell_1^2} {1\over
   \vec\ell_2^2}
 {1\over (\vec\ell_1+\vec\ell_2+\vec q)^2}{1\over
    (\vec\ell_1+\vec q)^2}\,.
\end{equation}

We now consider the  large mass expansion of the graph
\begin{equation}
 I_{\triangleright\triangleright (2)}(p_1,q)=
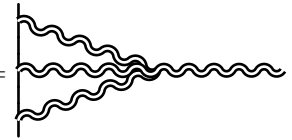
\begin{gathered}
 \begin{fmffile}{twolooptriangle4pt}
  \begin{fmfgraph}(100,50)
\fmfstraight
      \fmfleftn{i}{2}
      \fmfrightn{o}{1}
\fmf{plain,tension=10}{i1,ov1,ovph1,ovph2,ov3,ovph3,ovph4,ov2,i2}
\fmffreeze
\fmf{dbl_wiggly,tension=.3}{v3,ov1}
\fmf{dbl_wiggly,tension=.3}{v3,ov2}
\fmf{dbl_wiggly,tension=.3}{v3,ov3}
\fmf{dbl_wiggly,tension=1}{o1,v3}
\end{fmfgraph}
\end{fmffile}
\end{gathered}
\end{equation}
which to leading order reads
\begin{multline}
  I_{\triangleright\triangleright(d)}(p_1,q) = -{1\over 3}\int \prod_{i=1}^3 \left({d^4\ell_i\over (2\pi)^4}
  {1\over \ell_i^2+i\epsilon}\right)\cr
\times(2\pi)^3\delta^{(3)}(\ell_1+\ell_2+\ell_3+q)\cr
\times2\pi\delta(\ell_1^0+\ell_2^0+\ell_3^0)\Big(  {1\over
    2m_1\ell_1^0+i\epsilon}{1\over
   2m_1 \ell_2^0-i\epsilon}\cr+ {1\over
    2m_1\ell_3^0+i\epsilon}{1\over
    2m_1\ell_1^0-i\epsilon}+ {1\over
    2m_1\ell_3^0+i\epsilon}{1\over
    2m_1\ell_2^0-i\epsilon}\Big)\,,
\end{multline}
and evaluates to 
\begin{equation}\label{e:Tri2loop2}
   I_{\triangleright\triangleright(2)}(p_1,q) =  {1\over 12m_1^2}\!\! \int {d^3\vec \ell_1\over
     (2\pi)^3} \!\! {d^3\vec\ell_2\over(2\pi)^3}\, {1\over
     \vec\ell_1^2}{1\over \vec\ell_2^2} {1\over
     (\vec\ell_1+\vec\ell_2+\vec q)^2}\,.
\end{equation}
The expressions~\eqref{e:Tri2loop1} and~\eqref{e:Tri2loop2} are precisely the
coupling of three static sources to  a massless  tree amplitude 
\begin{equation}
 I_{\triangleright\triangleright(1)}(p_1,q), 
 I_{\triangleright\triangleright(2)}(p_1,q) \leftrightarrow\qquad
\begin{gathered}
\begin{fmffile}{Twoloopsources}
    \begin{fmfgraph*}(100,50)
\fmfstraight
       \fmfleftn{i}{3}
        \fmfrightn  {o}{1}
\fmf{dbl_wiggly}{i1,v1}
\fmf{dbl_wiggly}{i2,v1}
\fmf{dbl_wiggly}{i3,v1}
\fmf{dbl_wiggly,tension=3}{v1,o1}
\fmfblob{.5cm}{i1}
\fmfblob{.5cm}{i2}
\fmfblob{.5cm}{i3}
\fmfv{decor.shape=circle,decor.filled=30,decor.size=.3w}{v1}
\end{fmfgraph*}
\end{fmffile}
\end{gathered}
\end{equation}
A generalisation of the identity~\eqref{e:identity} implies that
the sum of all the permutation of  $n$ massless
propagators connected to a massive scalar line results in the
coupling of classical sources to multileg tree
amplitudes~\cite{Ludovic,Jules}. The same conclusion applies to massive particles with spin
as we will demonstrate elsewhere.


This analysis applies directly to the computation of an off-shell quantity such as the metric itself.
Consider the absorption of a graviton
\begin{multline}
  \langle p_2|T_{\mu\nu}|p_1\rangle={-i\over2m_1} \int {d^4\ell\over(2\pi)^4}{P_{\rho\sigma,\alpha\beta }\over \ell^2+i\epsilon}
{P_{\kappa\lambda,\gamma\delta}\over  (\ell+q)^2+i\epsilon}\cr
  {\tau_1^{\rho\sigma}(p_1-\ell,p_1)\tau_1^{\kappa\lambda}(p_1-\ell,p_2)
    \tau_{3\,\mu\nu}{}^{\alpha\beta,\gamma\delta}(\ell,\ell-q) 
 \over (\ell+p_1)^2-m_1^2+i\epsilon} \,,
\end{multline}
where $\tau_1$ is the vertex for the coupling of one graviton to a scalar
given in~\cite[eq.~(72)]{BjerrumBohr:2002ks}, $\tau_3$ is the three
graviton vertex given in~\cite[eq.~(73)]{BjerrumBohr:2002ks} and
$P_{\mu\nu,\rho\sigma}$ is the projection operator given in~\cite[eq.~(30)]{BjerrumBohr:2002ks}.
In the large $m$ limit $|q|/m\ll1$ projects the integral on the 00-component of the scalar vertex $\tau_1^{\mu\nu}(p_1-\ell,p_1)\simeq
i\kappa m_1^2 \delta^{\mu}_0 \delta^\nu_0$. Focusing on the 00
component we have in this limit~\cite{Jules}
\begin{multline}
   \langle p_2|T_{00}|p_1\rangle\simeq 4i\pi G m_1^3 \int {d^4\ell\over(2\pi)^4}
\left(\frac38{\vec q~}^2-\frac32 {\vec\ell~}^2\right)\cr
\times {1\over (\ell^2+i\epsilon)  ((\ell+q)^2+i\epsilon)( (\ell+p_1)^2-m_1^2+i\epsilon)}\cr
={3\kappa^2m^3\over128} |\vec q|\,,
\end{multline}
where we used the result of the previous section to evaluate the
triangle integral. This reproduces the classical first post-Newtonian contribution to the
00-component of the Schwarzschild metric evaluated in~\cite{BjerrumBohr:2002ks}. It  also immediately shows how to relate
a conventional Feynman-diagram evaluation with
the computation of Duff~\cite{Duff:1973zz} who derived such tree-like structures from classical sources.

\vspace{-.4cm}
\section{Scalar Interaction Potentials}\unskip

For the classical terms we need only the
graviton cuts, and instead of computing classes of diagrams, we apply
the unitarity method directly to get the on-shell scattering
amplitudes initiated in~\cite{Neill:2013wsa} and further developed in~\cite{Bjerrum-Bohr:2013bxa,Vaidya:2014kza}.
We first consider the scattering of two scalars of masses $m_1$ and $m_2$, respectively.
At one-loop order this entails a two-graviton cut of a massive scalar
four-point amplitude.  We have shown that classical terms arise from topologies with loops solely
entering as triangles that include the massive states. When we glue together the two on-shell scattering
amplitudes, we thus discard all terms that do not correspond to such topologies. Rational terms are not needed, as they correspond to
ultra-local terms of no relevance for the long-distance interaction potentials.

We first recall the classical tree-level result  from the one-graviton exchange

\begin{align}
\mathcal M_1&\!= \!\!\!
\begin{gathered}\begin{fmffile}{treegraph}\begin{fmfgraph}(60,30)
      \fmfleftn{i}{2}
      \fmfrightn{o}{2}
\fmf{plain}{o1,v3,o2}
\fmf{plain}{i1,v1,i2}
\fmf{dbl_wiggly,tension=.6,label{$q$}}{v1,v3}
\fmflabel{$p_1$}{i1}
\end{fmfgraph}
\end{fmffile}
\end{gathered}\cr
\label{e:tree}
&\!=\!-{16\pi G\over q^2}\!\! \left(m_1^2m_2^2\!-\!2(p_1\cdot p_2)^2\!-\!(p_1\cdot
  p_2) q^2\right),
\end{align}
where incoming momenta are $p_1$ and $p_2$ and $p_1^2={p_1'}^2=m_1^2$,
$p_2^2={p_2'}^2=m_2^2$ and the momentum transfer $q=p_1-p_1'=-p_2+p_2'$.

The two-graviton interaction is clearly a one-loop amplitude that can be
constructed using the on-shell unitarity
method~\cite{Neill:2013wsa,Bjerrum-Bohr:2013bxa,Vaidya:2014kza}. The previous analysis shows that the classical piece is
contained in the triangle graphs
\begin{align}\label{e:one}
&  \mathcal M_2 =\begin{gathered}\begin{fmffile}{triangleleft}
  \begin{fmfgraph}(60,30)
      \fmfleftn{i}{2}
      \fmfrightn{o}{2}
\fmf{plain}{i1,v1}
\fmf{plain}{i2,v2}
\fmf{plain}{o1,v3,o2}
\fmf{plain,tension=.1}{v1,v2}
\fmf{dbl_wiggly,tension=.3}{v3,v1}
\fmf{dbl_wiggly,tension=.3}{v3,v2}
\end{fmfgraph}
\end{fmffile}
\end{gathered}
+\begin{gathered}
\begin{fmffile}{triangleright}
  \begin{fmfgraph}(60,30)
      \fmfleftn{i}{2}
      \fmfrightn{o}{2}
\fmf{plain}{o1,v1}
\fmf{plain}{o2,v2}
\fmf{plain}{i1,v3,i2}
\fmf{plain,tension=.1}{v1,v2}
\fmf{dbl_wiggly,tension=.3}{v3,v1}
\fmf{dbl_wiggly,tension=.3}{v3,v2}
\end{fmfgraph}
\end{fmffile}\end{gathered}\\
\nonumber &\!\!=\! -i(8\pi G)^2\!\!\left(\! {c(m_1,m_2)    I_\triangleright(p_1,q)\over  \left(q^2-4m_1^2\!\right)^2}
\! +\!{ c(m_2,m_1)   I_\triangleright(p_2,-q) \over  \left(q^2-4m_2^2\right)^2} \right),
\end{align}
with for the interaction between two massive scalars 
\begin{multline}
c(m_1,m_2)=
(q^2)^5+ (q^2)^4 \left(6   p_1 \cdot p_2-10m_1^2\right)\cr
+ (q^2)^3 \left(12 (p_1 \cdot p_2)^2-60 m_1^2 p_1\cdot p_2-2m_1^2m_2^2+30 m_1^4\right)\cr
-(q^2)^2 \left(120m_1^2(p_1\cdot p_2)^2-180 m_1^4 p_1\cdot
  p_2-20m_1^4 m_2^2+20m_1^6\right)\cr
+q^2\left(360m_1^4(p_1\cdot p_2)^2-120 m_1^6 p_1\cdot p_2-4m_1^6
  (m_1^2+15m_2^2)\right)\cr
+48m_1^8m_2^2-240m_1^6 (p_1\cdot p_2)^2\,.
\end{multline}

At leading order in $q^2$, using the result~\eqref{e:Triresult} the
two gravitons exchange simplifies to just, in agreement
with~\cite[eq~(3.26)]{Guevara} and~\cite{Neill:2013wsa,Cachazo:2017jef},
\begin{equation}\label{e:one2}
  \mathcal M_2 = {6\pi^2 G^2\over |\vec q|}  (m_1+m_2)(5(p_1\cdot
  p_2)^2-m_1^2m_2^2)+O(|\vec q|) \,.
\end{equation}

Note the systematics of this expansion. The Einstein metric is expanded perturbatively, and all physical momenta are provided at infinity. Contractions
of momenta are performed with respect to the flat-space Minkowski metric only, and no reference is made to space-time coordinates. This is a
gauge invariant expression for the classical scattering amplitude in a
plane-wave basis that
is independent of coordinate choices (and gauge choices). 
To derive a classical
non-relativistic potential, we need to choose coordinates: We Fourier transform the gauge invariant momentum-space scattering amplitude. This introduces coordinate dependence even in theories such as Quantum Electrodynamics. Moreover, just as in Quantum Electrodynamics, we must also be careful in keeping sub-leading terms of this Fourier transform and thus 
expand in $q^0$ consistently.  This forces us to keep velocity-dependent terms in the energy that are of the same order as the naively defined static 
potential. One easily checks that the  overall sign of the 
amplitudes in~\eqref{e:tree} and~\eqref{e:one2} are precisely the ones
required for an attractive force.

The result of this procedure has been well documented elsewhere, starting with the pioneering observations of Iwasaki~\cite{Iwasaki:1971vb}, and later
reproduced in different coordinates
in~\cite{Holstein:2008sx,Neill:2013wsa}. Although we are unable to
reproduce the individual contributions in~\cite[eqs.~(A.1.4)--(A.1.6)]{Iwasaki:1971vb} our final result for the interaction energy is
to this order 
\begin{align}
H &= \frac{\vec{p}_1^2}{2m_1} + \frac{\vec{p}_2^2}{2m_2} - \frac{\vec{p}_1^4}{8m_1^3} - \frac{\vec{p}_2^4}{8m_2^3} \\
&-\frac{Gm_1m_2}{r} - \frac{G^2m_1m_2(m_1+m_2)}{2r^2} \cr
\nonumber &- \frac{Gm_1m_2}{2r}\left(\frac{3\vec{p}_1^2}{m_1^2} + \frac{3\vec{p}_2^2}{m_2^2} - \frac{7\vec{p}_1\cdot\vec{p}_2}{m_1m_2}
- \frac{(\vec{p}_1\cdot\vec{r})(\vec{p}_2\cdot\vec{r})}{m_1m_2r^2}\right),
\end{align}
which precisely leads to the celebrated Einstein-Infeld-Hoffmann equations of motion. It is crucial to correctly perform the subtraction of the iterated
tree-level Born term in order to achieve this.

\section{The post-Minkowskian expansion}

 The scattering problem of general relativity can be treated in a fully relativistic manner, without a truncated expansion in velocities.
To this end, we consider here the full relativistic scattering amplitude and expand in Newton's constant $G$ only. For the scalar-scalar case we thus return to
the complete classical one-loop result~(\ref{e:one}). The conventional Born-expansion expression that is used to derive the quantum mechanical cross
section is not appropriate here, even if we keep only the classical part of the amplitude. That expression for the cross section is based on incoming plane waves, and will not not match the corresponding
classical cross section beyond the leading tree-level term. In fact, even the classical cross section is unlikely to be of any interest observationally. 
So a more
meaningful approach is to use the classical scattering amplitude to compute the scattering angle of two masses colliding with a given impact parameter $b$.

We use the eikonal approach to derive the relationship between small scattering angle $\theta$ and impact parameter $b$. Generalizing the analysis of
ref.~\cite{Kabat:1992tb} and~\cite{Akhoury:2013yua} (see the Appendix for some details
   on the eikonal method to one-loop order) to the case of two scalars of masses $m_1$ and $m_2$, we focus on the high-energy regime 
$s,t$ large and $t/s$ small. Note that in addition to expanding in $G$, we are also expanding
the full result~(\ref{e:one}) 
in $q^2$, and truncating already at next-to-leading order.
We go to the center of mass frame and define $p \equiv |\vec{p_1}| = |\vec{p_2}|$. The impact parameter is defined by a two-dimensional vector 
$\vec{b}$ in the plane of scattering orthogonal to $\vec{p}_1 = -\vec{p}_2$, with $b \equiv |\vec{b}|$.
In the eikonal limit we find the exponentiated relationship between the scattering amplitude 
\begin{equation}\label{e:Mb}
M(\vec{b}) ~\equiv~ \int d^2\vec{q} e^{-i\vec{q}\cdot\vec{b}}M(\vec{q})\,, 
\end{equation}
and scattering function $\chi(b)$ to be
\begin{equation}\label{e:Mexp}
M(\vec{b}) ~=~ 4p(E_1+E_2) (e^{i\chi(\vec{b})}-1) ~.
\end{equation}
In order to compare with the first computation of post-Minkowskian scattering to order $G^2$~\cite{Westpfahl}, we introduce new kinematical variables 
$M^2 \equiv s$,
$\hat{M}^2 \equiv M^2 - m_1^2 - m_2^2$. We go to the center of mass frame where $p^2 = 
(\hat M^4-4m_1^2m_2^2)/4M^2$. In terms of the scattering angle
$\theta$ we have $t \equiv q^2 = [(\hat
M^4-4m_1^2m_2^2)\sin^2(\theta/2)]/M^2$, and $4p(E_1+E_2)=2\sqrt{\hat M^4-4m_1^2m_2^2}$. Keeping,
consistently, only the leading order in $q^2$ of the one-loop amplitude (\ref{e:Mb}), we obtain
\begin{equation}
 2\sin(\theta/2)\!=\! {-2M\over \sqrt{\hat M^4-4m_1^2m_2^2}}{\partial\over\partial b}\left(\chi_1(b)
    +\chi_2(b)\right) ,
\end{equation}
where $\chi_1(b)$ and $\chi_2(b)$ are the tree-level and one-loop
scattering functions given respectively by the Fourier
transform of the scattering amplitudes
\begin{equation}\label{e:chi1}
  \chi_i(b)= {1\over 2\sqrt{\hat M^4-4m_1^2m_2^2}} \int {d^2\vec q\over
    (2\pi)^2}\, e^{-i\vec q\cdot \vec b} \mathcal M_i(\vec q) \,. 
\end{equation}
At leading order in $q^2$ the tree-level and one-loop amplitudes
in~\eqref{e:tree} and~\eqref{e:one2} read
\begin{align}
    \mathcal M_1(\vec q)&= {8 \pi G\over {\vec q}^2}\, (\hat
                             M^4-2m_1^2m_2^2)\,,\cr
  \mathcal M_2(\vec q)&= {3\pi^2 G^2\over
      2|\vec q|}\,(m_1+m_2) (5\hat M^4-4m_1^2m_2^2)\,,
\end{align}
where higher order terms in $q^2$ correspond in position space to quantum corrections.
Only the triangle contribution contribute to the one-loop 
scattering function because the contributions from the boxes and
cross-boxes contributed to the exponentiation of the tree-level
amplitude~\cite{Akhoury:2013yua,Bjerrum-Bohr:2014zsa}.
The  Fourier transform around two dimensions is computed using 
\begin{multline}
\mu^{2-D}  \int {d^Dq\over(2\pi)^D} e^{-i\vec q\cdot \vec b}|\vec
q|^\alpha\cr= {(2\pi\mu)^{2-D}\over 4\pi}\left(2\over b\right)^{\alpha+D}
{\Gamma\left(\alpha+D\over2\right)\over \Gamma\left(2-\alpha-D\over2\right)}\,.
\end{multline}
The scattering functions then read
\begin{align}\label{e:chi12result}
\chi_1(b)&= 2G\, {\hat M^4-2m_1^2m_2^2\over \sqrt{\hat M^4-4m_1^2m_2^2}}\,
  \left({1\over d-2}-\log(\pi\mu b)-\gamma_E\right)\,,\cr
  \chi_2(b)&= {3\pi G^2\over 8\sqrt{\hat M^4-4m_1^2m_2^2}}
             {m_1+m_2\over b} (5\hat M^4-4m_1^2m_2^2)\,,
\end{align}
where $\mu$ is a regularisation scale.
The scattering angle to this order reads
\begin{multline}
  2\sin\left(\theta\over2\right)={4GM\over b} \Big( {\hat M^4-2m_1^2m_2^2\over \hat
  M^4-4m_1^2m_2^2}\cr
+  {3\pi\over 16}{G(m_1+m_2)\over b}{5\hat M^4-4m_1^2m_2^2\over
  \hat M^4-4m_1^2m_2^2} \Big).
\end{multline}
This result agrees with the expression found by Westpfahl~\cite{Westpfahl} who explicitly solved the Einstein equations to this order in $G$ and in the same 
limit of small scattering angle. We find the present approach to be
superior in efficiency, and very easily generalizable to higher orders
in $G$.

Taking the massless limit $m_2 = 0$ and approximating $2\sin(\theta/2) \simeq \theta$, we recover the classical bending angle of light
$\theta = \frac{4Gm_1}{b} + \frac{15\pi}{4}\frac{G^2m_1^2}{b^2},$
including its first non-trivial correction in $G$, in agreement with~\cite[\S101]{Landau:1982dva}.
We have additionally computed the full expression for the classical part of the scalar-fermion 
(spin 1/2) amplitude up to and including one-loop order, but do not display the results here for lack of space. We stress that the small-angle scattering formula derived above is based on only a small amount of the information contained in the full one-loop scattering amplitude~(\ref{e:one}).

\section{Light-by-light scattering in general relativity}

 Photon-photon scattering is particularly interesting, as our analysis will show how to derive an exact result in general relativity. As explained above, classical
contributions from loop diagrams require the presence of massive triangles in the loops. For photon-photon scattering there are no such contributions to any
order in the expansion, and we
conclude that photon-photon scattering in general relativity is
tree-level exact

\begin{align}
 & M_{\gamma\gamma}=
\begin{gathered}\begin{fmffile}{treegraphphotons}
\begin{fmfgraph}(60,30)
      \fmfleftn{i}{2}
      \fmfrightn{o}{2}
\fmf{photon}{o1,v3,o2}
\fmf{photon}{i1,v1,i2}
\fmf{dbl_wiggly,tension=.6}{v1,v3}
\end{fmfgraph}
\end{fmffile}\end{gathered}
+\begin{gathered}
\begin{fmffile}{treegraphphotont}
\begin{fmfgraph}(60,30)
      \fmfleftn{i}{2}
      \fmfrightn{o}{2}
\fmf{photon,tension=5}{i1,v1,o1}
\fmf{photon,tension=5}{i2,v2,o2}
\fmf{dbl_wiggly}{v1,v2}
\end{fmfgraph}
\end{fmffile}\end{gathered}
+\begin{gathered}
\begin{fmffile}{treegraphphotonu}
\begin{fmfgraph}(60,30)
      \fmfleftn{i}{2}
      \fmfrightn{o}{2}
\fmf{photon,tension=2}{i1,v1}
\fmf{phantom,tension=2}{v1,o1}
\fmf{photon,tension=2}{i2,v2}
\fmf{phantom,tension=2}{v2,o2}
\fmf{dbl_wiggly,tension=.2}{v1,v2}
\fmf{photon,tension=.2}{v2,o1}
\fmf{photon,tension=.2,rubout}{v1,o2}
\end{fmfgraph}
\end{fmffile}\end{gathered}
\\
&-8\pi G{ \,2\tr(f_1f_2f_3f_4)+2\tr(f_1f_3f_4f_2)-\tr(f_1f_2)\tr(f_3f_4)\over  (p_1-p_2)^2}\cr
&-8\pi G\,
  {2\tr(f_1f_4f_3f_2)+2\tr(f_1f_3f_2f_4)-\tr(f_1f_4)\tr(f_2f_3)\over  (p_1+p_4)^2} \cr
&-8\pi G\,
\nonumber {2\tr(f_1f_3f_4f_2)+2\tr(f_1f_3f_2f_4)-\tr(f_1f_3)\tr(f_2f_4)\over (p_1-p_3)^2}\,,
\end{align}
where  the traces are evaluated over the Lorentz indices and
$f_i^{\mu\nu}=\epsilon_i^\mu p_i^\nu-\epsilon_i^\nu p_i^\mu$ are the
field-strength of the photon fields.
When considering polarised photons, it is immediate to check that this
amplitude is non-vanishing only for scattering of photons of opposite
helicity as no force is expected between photons of the same
helicity.  Similarly the force between parallel photons
vanishes~\cite{gupta}.

\vspace{-.6cm}
\section{Conclusion}
\vspace{-.3cm}

We have explicitly shown how loops of
the Feynman diagram expansion become equal to the tree-like structures
coupled to classical sources thus demystifying the
appearance of loop diagrams in classical gravity, and, at the same time, linking the source-based method directly to conventional 
Feynman diagrams. 
Interestingly, the manner in which the $\ell^0$-integrations conspire to leave tree-like structures from loops of triangle graphs also
forms the precise bridge to classical general relativity computations based on the world-line formulation (see, $e.g.$,
\cite{Muzinich:1995uj,Goldberger:2004jt,Gilmore:2008gq,Foffa:2011ub,Blanchet:2013haa,Foffa:2016rgu,Porto:2016pyg}).

Enormous simplifications occur when computing
what corresponds to on-shell quantities, based on the unitarity method~\cite{Bern:1994zx}.
Non-analytic terms~\cite{Donoghue:1993eb} involving powers of $m/\sqrt{-q^2}$  
produce the long-distance classical contributions from the loops. By the rules of unitarity
cuts, we can reconstruct these non-analytic pieces by gluing tree-level amplitudes together while summing over physical
states of the graviton legs only~\cite{Neill:2013wsa,Bjerrum-Bohr:2013bxa,Vaidya:2014kza,Cachazo:2017jef,Guevara,Bjerrum-Bohr:2014zsa}. 

Scalar interaction potentials form the backbone of gravitational wave
computations for binary mergers.  The fact that the unitarity method
provides these results straightforwardly provides hope that this is
the beginning of a new approach to both post-Newtonian and post-Minkowskian calculations in
general relativity, including those relevant for the physics of gravitational waves.  
Since the method applies to the general
effective field theory of gravity, this opens up a way to
constrain  terms beyond the Einstein-Hilbert action that may affect the
observational signal of gravitational waves.

\vspace{-.2cm}
\section{Acknowledgments}\unskip
We thank John Joseph Carrasco, Thibault Damour, Barak Kol, Rafael Porto, Dennis R\"atzel and
Ira Rothstein for useful comments on the manuscript.
N.E.J.B and P.H.D. are supported in part by the Danish National Research Foundation (DNRF91). P.V. is partially supported by ``Amplitudes'' ANR-17-
CE31-0001-01, and Laboratory of Mirror
Symmetry NRU HSE, RF Government grant, ag. N$^\circ$
14.641.31.0001. G.F. are supported by the ERC STG grant 639220.

\appendix
\vspace{-.4cm}
\section{Appendix: The finite massive triangle integral}\label{app:triangle}
\vspace{-.4cm}

We evaluate the ultraviolet and infrared finite triangle integral for
$p^2>0$ and $q^2<0$
\begin{equation}
\!  I_\triangleright(p,q) \!=\!\!\int\!\! {d^4\ell\over (2\pi)^4} {1\over (\ell^2\!+\!i\epsilon) ((\ell\!+\!q)^2\!+\!i\epsilon) ((\ell\!+\!p)^2\!-\!p^2\!+\!i\epsilon)}\,.
\end{equation}
The parametric representation of this integral is~\cite{Vanhove:2014wqa}
\begin{equation}
\!\!\!\!  I_\triangleright(p_1,q)= {i\over 16\pi^2}\int_{x_1\geq0\atop x_2\geq0}
  {dx_1dx_2\over (1+x_1+x_2)(q^2x_1x_2-p^2)}\,.
\end{equation}
Integrating over $x_1$ leads to 
\begin{equation}
    I_\triangleright(p_1,q)=i
    \int_0^\infty {\log(1+x_2)+\log(x_2)+\log\left(-{q^2\over p^2}\right)\over
      16\pi^2 (p^2+q^2 x_2(1+x_2))}\,dx_2\,.
\end{equation}
We are interested in the case  $q^2/p^2<0$, and we setting 
$\rho_+ \equiv -\frac12+\frac12\sqrt{1-4p^2/q^2}$ and
$\rho_-\equiv -\frac12-\frac12\sqrt{1-4p^2/q^2}$. 
It is immediate to evaluate the integration over $x_2$
\begin{multline}
  I_\triangleright(p_1,q)=i {6\zeta(2)+ \log\left(-\rho_+\rho_-\right)
    \log\left(-{\rho_+\over \rho_-}\right)\over 16\pi^2 q^2 \sqrt{1-4p^2/q^2}}\cr
+i{\textrm{Li}_2\left(2,
      -{1\over \rho_+}\right)-\textrm{Li}_2\left(2, -{1\over
        \rho_-}\right)\over 16\pi^2 q^2 \sqrt{1-4p^2/q^2}}\,,
\end{multline}
where we have introduced the dilogarithm function (see~\cite{Zagier} for a
definition and properties)
\begin{equation}
  \textrm{Li}_2(z) \equiv -\int_0^z \log(1-t) {dt\over t}\,,     \qquad
  \textrm{for}\qquad z\in\mathbb C\,.
\end{equation}
For small $|q^2/p^2|\ll 1$ we find
\begin{equation}
  I_\triangleright(p_1,q)={i\over32 q^2} \sqrt{-{q^2\over p^2} }+O(q^2/p^2)\,.
\end{equation}

\section{Appendix: The scattering angle from the eikonal}

Our derivation of the classical scattering angle of two-body dynamics
in the post-Minkowskian expansion of general relativity  relied crucially
on the exponentiation of the scattering function for the appropriate kinematics. This kinematic regime is that of the eikonal, 
covariantly described by $s$ and $t$, with $t/s \ll 1$. It
is very conveniently, and symmetrically, analysed in the centre of mass frame. Here, momentum exchange across the massive
lines  $\vec{q}$ is treated as small. When we go from the plane-wave basis to describing point-particle scattering by means of a
Fourier transform to impact parameter space this is equivalent to keeping only the classical part of the scattering function. Higher orders
in $\vec{q}$ will correspond to quantum corrections, of no interest to us at this stage. Indeed, all multiple exchanges of gravitons of
arbitrarily high loop order in the eikonal limit are not affected by the ultraviolet divergences of the quantum theory of gravity~\cite{Kabat:1992tb,Luna:2016idw}, as expected for 
integrals that should only be related to classical general relativity. 

The main expressions for the eikonal limit have already been provided in the companion paper.
In this supplementary material we provide a detailed derivation of these formulae. The basic ingredients were described in the seminal paper~\cite{Kabat:1992tb}. In ref.~\cite{Akhoury:2013yua} this was extended in a highly non-trivial manner to the next eikonal order for the 
special situation of a very light particle scattering gravitationally off a heavy mass. Our remaining task is therefore to generalise the analysis of 
ref.~\cite{Akhoury:2013yua} to the general case of scattering of two
particles of masses $m_1$ and $m_2$.  We use the notation of
refs.~\cite{Akhoury:2013yua,Bjerrum-Bohr:2016hpa}. A somewhat similar analysis from
the viewpoint of string theory (and supergravity) has recently appeared in ref.~\cite{Collado:2018isu}.
  
Consider the scattering in the centre of mass frame. We go from the plane-wave basis of scattering to classical point-particle 
scattering by fixing an impact parameter $\vec{b}$ at far infinity, where space-time is Minkowskian. This is the two-dimensional vector
in the scattering plane that provides the initial data needed to
describe the classical trajectories of the point-like particles.

\section{The eikonal phase}

The one-graviton exchange is given by %
\begin{equation}\label{e:Onegrav}
    \mathcal M_1(\vec q)= {8\pi G\over {\vec q}^2}\, (\hat M^4-2m_1^2m_2^2)
  \end{equation}
 as a function of $q\equiv p_1-p_1'=-p_2+p_2'$.  
The symmetrized iteration of this amplitude to one-loop order gives, in the small-$\ell$ limit,
\begin{multline}
{\cal M}_{1}^{(2)}(\vec q) = \int
{d^4\ell\over (2\pi)^4} { (8\pi G  (\hat M^4-2m_1^2m_2^2))^2 \over
  (\ell^2+i\epsilon) ((q+\ell)^2+i\epsilon)} \cr
\times\frac12\,\Big({i\over 2\ell\cdot p_1+i\epsilon}{i\over
    -2\ell\cdot p_2+i\epsilon}
  +{i\over 2\ell\cdot p_1+i\epsilon}{i\over
    2\ell\cdot p_2'+i\epsilon}\cr
+  {i\over -2\ell\cdot p_1'+i\epsilon}{i\over
  -2\ell\cdot p_2+i\epsilon}
+{i\over -2\ell\cdot p_1'+i\epsilon}{i\over
    2\ell\cdot p_2'+i\epsilon}
\Big)\,.
\end{multline}
Following~\cite{Kabat:1992tb} we introduce the Fourier transform of
the propagator
\begin{equation}
  {1\over k^2+i\epsilon}= \int d^4x e^{-ik\cdot x} \Delta (x)\,  ~,
\end{equation}
allowing us to write
\begin{equation}
  \mathcal M_1^{(2)}(\vec q)=8\pi G  (\hat M^4-2m_1^2m_2^2) \int d^4x
  e^{-iq\cdot b} \Delta (b) \, {\chi_1  (b)  \over 2}
\end{equation}
where 
\begin{multline}\label{e:chiDef}
  \chi_1(b)= \int {d^4\ell\over
    (2\pi)^4}  e^{-i\ell\cdot b} { -8\pi G  (\hat M^4-2m_1^2m_2^2)\over \ell^2+i\epsilon} \cr
\times\Big({1\over 2\ell\cdot p_1+i\epsilon}{1\over
    -2\ell\cdot p_2+i\epsilon}
  +{1\over 2\ell\cdot p_1+i\epsilon}{1\over
    -2\ell\cdot p_2'+i\epsilon}\cr
+  {1\over 2\ell\cdot p_1'+i\epsilon}{1\over
  -2\ell\cdot p_2+i\epsilon}
+{1\over 2\ell\cdot p_1'+i\epsilon}{1\over
    -2\ell\cdot p_2'+i\epsilon}
\Big)\,,
\end{multline}
is the symmetrized sum over the permutations of the box and crossed box.
In the eikonal limit of, in first approximation, neglected recoil, $p_1\simeq p_1'$ and $p_2\simeq p_2'$ so that
$\chi_1$ becomes
\begin{multline}
  \chi_1(b)= \int {d^4\ell\over
    (2\pi)^4}  e^{-i\ell\cdot b} { -8\pi G  (\hat M^4-2m_1^2m_2^2)\over \ell^2+i\epsilon} \cr
\times \Big({1\over 2p_1\cdot \ell+i\epsilon} -{1\over 2p_1\cdot
  \ell-i\epsilon} \Big)
\Big({1\over 2p_2\cdot \ell+i\epsilon} -{1\over 2p_2\cdot
  \ell-i\epsilon} \Big)
\,.
\end{multline}
Using the identity
\begin{equation}
  \lim_{\epsilon\to 0}\left(  {1\over x+i\epsilon}- {1\over
    x-i\epsilon}\right)=-2i\pi \delta(x)
\end{equation}
we obtain
\begin{multline}
  \chi_1(b) =8\pi^3 G (\hat M^4\!-\!2m_1^2m_2^2) \cr
  \times \int {d^4\ell\over
    (2\pi)^4}  e^{-i\ell\cdot b} {(2\pi)^2\delta(2p_1\cdot \ell)\delta(2p_2\cdot \ell) \over
    \ell^2+i\epsilon} 
                       \,.
\end{multline}

We choose kinematics so that $p_1 = (E_1, 0, 0, p)$ and $p_2 = (E_2, 0, 0 , -p)$, Integration
over the longitudinal components $\ell_0$  and $\ell_3$ then lead to the Jacobian factor
\begin{equation}
  \label{eq:5}
  \int {d^4\ell\over
    (2\pi)^2}  \delta(2p_1\cdot \ell)\delta(2p_2\cdot \ell)= {1\over 4p(E_1+E_2)}\int {d^2\vec\ell_\perp\over    (2\pi)^2}
\end{equation}
with $p=|\vec p_1|=|\vec p_2|$. Thus,
\begin{equation}\label{e:chi1final}
    \chi_1( b) = {8\pi G  (\hat M^4-2m_1^2m_2^2)\over 4p(E_1+E_2)}\int {d^2\vec\ell_\perp\over
    (2\pi)^2}  {e^{-i\vec\ell_\perp\cdot \vec b_\perp} \over
    \vec\ell_\perp^2+i\epsilon}
\,,
\end{equation}

This shows that the $\chi_1(b)$ is essentially the two-dimensional Fourier transform of the
tree-level one-graviton exchange as given in equation~(24) of the
main text.

We evaluate the $n$th iteration of the tree-level one-graviton
exchange
\begin{equation}
  \mathcal M_1^{(n+1)}(q)=8\pi G (\hat M^4-2 m_1^2m_2^2) \int d^4x
  e^{-iq\cdot b} \Delta (b) \, {\chi_1^{(n+1)}(b)\over (n+1)!}
\end{equation}
where $\chi_1^{(n+1)}(b)$, which generalises~(\ref{e:chiDef}), is the $n$th iteration of the one-graviton
exchange summing all permutations of ladder diagrams.  Using the algebraic identity provided in~\cite[Chap.~6]{Peskin:1995ev}
the symmetrized exchange of $n$ iterations of the one-graviton amplitude in this eikonal limit then gives
\begin{align}
 \chi_1^{(n+1)}(b)&={(8\pi G  (\hat M^4-2m_1^2m_2^2))^n
   (4p(E_1+E_2))^n}\cr
& \times\int \prod_{r=1}^{n}  {d^2\vec\ell_\perp^{r}\over
    (2\pi)^2}  {e^{-i\vec\ell_\perp^r \cdot \vec b} \over
    ( \vec\ell_\perp^r)^2+i\epsilon}\cr
  &=(\chi_1(b))^n\,.
\end{align}
Defining $\mathcal M_1^{(1)}(q) \equiv \mathcal M_1(q)$,
the sum over all iterations of the one-graviton exchange
\begin{equation}
  \mathcal M_1^{\textrm{sum}}(q)  = \sum_{n\geq1} \mathcal
  M_1^{(n)}(q) 
\end{equation}
then leads to
\begin{multline}
  \mathcal M_1^{\textrm{sum}}(q)  = 8\pi G (\hat M^4-2m_1^2m_2^2)\cr\times\int
  d^4b e^{-iq\cdot b} \Delta (b) \,  {e^{i\chi_1(b)}-1\over \chi_1(b)}\,.
\end{multline}
Using the expression~(\ref{e:chi1final}) for $\chi_1(b)$ and noting
that 
\begin{align}
  \int d^2 x_\parallel \Delta(x)&= \int d^2x_\parallel \int {d^4k\over
    (2\pi)^4} \, {e^{i (k_\perp\cdot x_\perp+k_\parallel \cdot
                                  x_\parallel)}\over k_\perp^2+k_\parallel^2+i\epsilon}\cr
&=         \int {d^2k_\perp\over
    (2\pi)^2} \, {e^{i k_\perp\cdot x_\perp}\over k_\perp^2+i\epsilon} ~,                       
\end{align}
it is then
immediately clear that in the eikonal limit the expression reduces to~\cite{Kabat:1992tb}
\begin{equation}
\chi_1(b)= {8\pi G (\hat M^4-2m_1^2m_2^2)\over4p (E_1+E_2)} \, \int
d^2b_\parallel \Delta(b)\,
\end{equation}
Therefore, the sum over all iterations of the one-graviton
exchange~(\ref{e:Onegrav}) reads
\begin{equation}\label{e:sumfinal}
  \mathcal M_1^{\textrm{sum}}(q)  = 4p(E_1+E_2)\,\int
  d^2b_\perp e^{-iq\cdot b_\perp} \,  \left(e^{i\chi_1(b)}-1\right)\,.
\end{equation}
as provided in equation~\eqref{e:Mexp} in the main text.

\section{The next-to-eikonal contribution}
The inclusion of the one-loop two-graviton exchange
\begin{equation}
  \mathcal M_2(\vec q)  = {3\pi^2G^2\over 3|\vec q|}(m_1+m_2)(5\hat M^4-4m_1^2m_2^2)\,.
\end{equation}
is far more involved, but the result of the analysis boils down to
a next-to-eikonal correction that also exponentiates, in exactly the same manner as shown above. This has been
demonstrated in detail in~\cite{Akhoury:2013yua} for the case of a nearly massless particle scattering off a heavy particle.
The generalization of that analysis to the scattering of two masses $m_1$ and $m_2$ immediately reveal that the
pattern shown above for tree-level exchanges are reproduced. In particular, the crucial Jacobian factor that arises
from the integration over the longitudinal components of the loop momentum is unchanged.
The summation of the iterated one-loop exchange therefore leads to
\begin{multline}
  \mathcal M_1^{\textrm{sum}}(q)  +\mathcal M_2^{\textrm{sum}}(q) \cr= 4p(E_1+E_2)\,\int
  d^2b_\perp e^{-iq\cdot b_\perp} \,  \left(e^{i(\chi_1(b)+\chi_2(b))}-1\right)\,.
\end{multline}
where the exponentiation of $\chi_2(b)$ arises from iterated exchanges combining the
one-graviton and the two-graviton amplitudes (see eq.~(5.22) of~\cite{Bjerrum-Bohr:2016hpa}).
The next-to-eikonal phase contribution thus reads
\begin{equation}
  \chi_2(b)= {1\over 4p(E_1+E_2)} \, \int
  {d^2\vec\ell_\perp\over(2\pi)^2} {e^{-i\vec\ell_\perp\cdot\vec
      b_\perp}} \mathcal M_2(\vec q)\,,
\end{equation}
and
\begin{multline}
  \mathcal M_1^{\textrm{sum}}(q)  +\mathcal M_2^{\textrm{sum}}(q) \cr
  = 4p(E_1+E_2)\,\int
  d^2b_\perp e^{-iq\cdot b_\perp} \,  \left(e^{i(\chi_1(b)+\chi_2(b))}-1\right)\,.
\end{multline}
The remaining two-dimensional integrals needed to evaluate the
scattering functions $\chi_1(b)$ and $\chi_2(b)$ can now be performed.
The  Fourier transform around two dimensions is computed using 
\begin{multline}
\mu^{2-D}  \int {d^Dq\over(2\pi)^D} e^{-i\vec q\cdot \vec b}|\vec
q|^\alpha\cr= {(2\pi\mu)^{2-D}\over 4\pi}\left(2\over b\right)^{\alpha+D}
{\Gamma\left(\alpha+D\over2\right)\over \Gamma\left(2-\alpha-D\over2\right)}\,.
\end{multline}
The leading term of $\chi_1(b)$ has the usual logarithmic divergence of a 
two-dimensional massless propagator using that around $D=2$ dimensions
\begin{equation}
  \mu^{2-D}\int\frac{d^Dq}{(2\pi)^D}{e^{-\vec{q}\cdot\vec{b}}\over \vec{q}^2} = \frac{1}{2\pi(D-2)} -{1\over2\pi} \log(b\mu\pi)-{\gamma_E\over2\pi}
\end{equation}
but as expected the divergence plays no role as we need only the derivative with respect to $b$.
The other integral is elementary,
\begin{equation}
\int\frac{d^2q}{(2\pi)^2}{e^{-\vec{q}\cdot\vec{b}}\over |\vec{q}|} = \frac{1}{2\pi b} ~.
\end{equation}
In this way we finally arrive at eq.~\eqref{e:chi12result} of the main text.

We finally note that since submission of the paper another post-Minkowskian approach to the 
two-body interaction potentials has appeared in~\cite{Cheung:2018wkq}. 


\end{document}